# On the imaging of electron transport in semiconductor quantum structures by scanning-gate microscopy: successes and limitation.

H. Sellier[1], B. Hackens[2,§], M.G. Pala[3], F. Martins[2], S. Baltazar[3], X. Wallart[4], L. Desplanque[4], V. Bayot[1,2], S. Huant[1,§]

[1] Institut Néel, CNRS and Université Joseph Fourier, BP 166, 38042 Grenoble, France

[2] IMCN, Pôle NAPS, Université Catholique de Louvain, B-1348 Louvain-la-Neuve, Belgium

[3] IMEP-LAHC, UMR 5130, CNRS/INPG/UJF/UdS, Minatec, Grenoble INP, Grenoble, France

[4] IEMN, UMR CNRS 8520, UST Lille, BP 60069, 59652 Villeneuve d'Ascq, France

[§] Authors to whom correspondence should be addressed

E-mail : benoit.hackens@uclouvain.be

serge.huant@grenoble.cnrs.fr

**Abstract.** This paper presents a brief review of scanning-gate microscopy applied to the imaging of electron transport in buried semiconductor quantum structures. After an introduction to the technique and to some of its practical issues, we summarise a selection of its successful achievements found in the literature, including our own research. The latter focuses on the imaging of GaInAs-based quantum rings both in the low magnetic field Aharonov-Bohm regime and in the high-field quantum Hall regime. Based on our own experience, we then discuss in detail some of the limitations of scanning-gate microscopy. These include possible tip induced artefacts, effects of a large bias applied to the scanning tip, as well as consequences of unwanted charge traps on the conductance maps. We emphasize how special care must be paid in interpreting these scanning-gate images.

# 1. Introduction

Scanning-gate microscopy (SGM) uses the electrically polarized tip of a low-temperature atomic-force microscope (AFM) to scan above a semiconductor device while the conductance changes due to the tip perturbation are simultaneously mapped in real space. Since its introduction in the late nineties [1-3], SGM has proven powerful to unravel the local details of electron behaviour inside modulation-doped nanostructures whose active electron systems are, in contrast to surface electron-systems, not accessible to scanning-tunnelling microscopy (STM) because they are buried too deep below the free surface. In this paper, we first describe the technique and some technical issues related to it (section 2). Then, we give in section 3 a review of some of its instructive achievements found in the literature, with a special emphasis on our own research that deals with the imaging of electron transport across GaInAs quantum rings both in the (low-field) Aharonov-Bohm (AB) regime and in the (high-field) quantum Hall regime. Starting from our own experience, we discuss in section 4 the main limitations of the SGM technique that may be encountered and emphasize where caution must be paid either to operate SGM or to interpret the images that it produces.

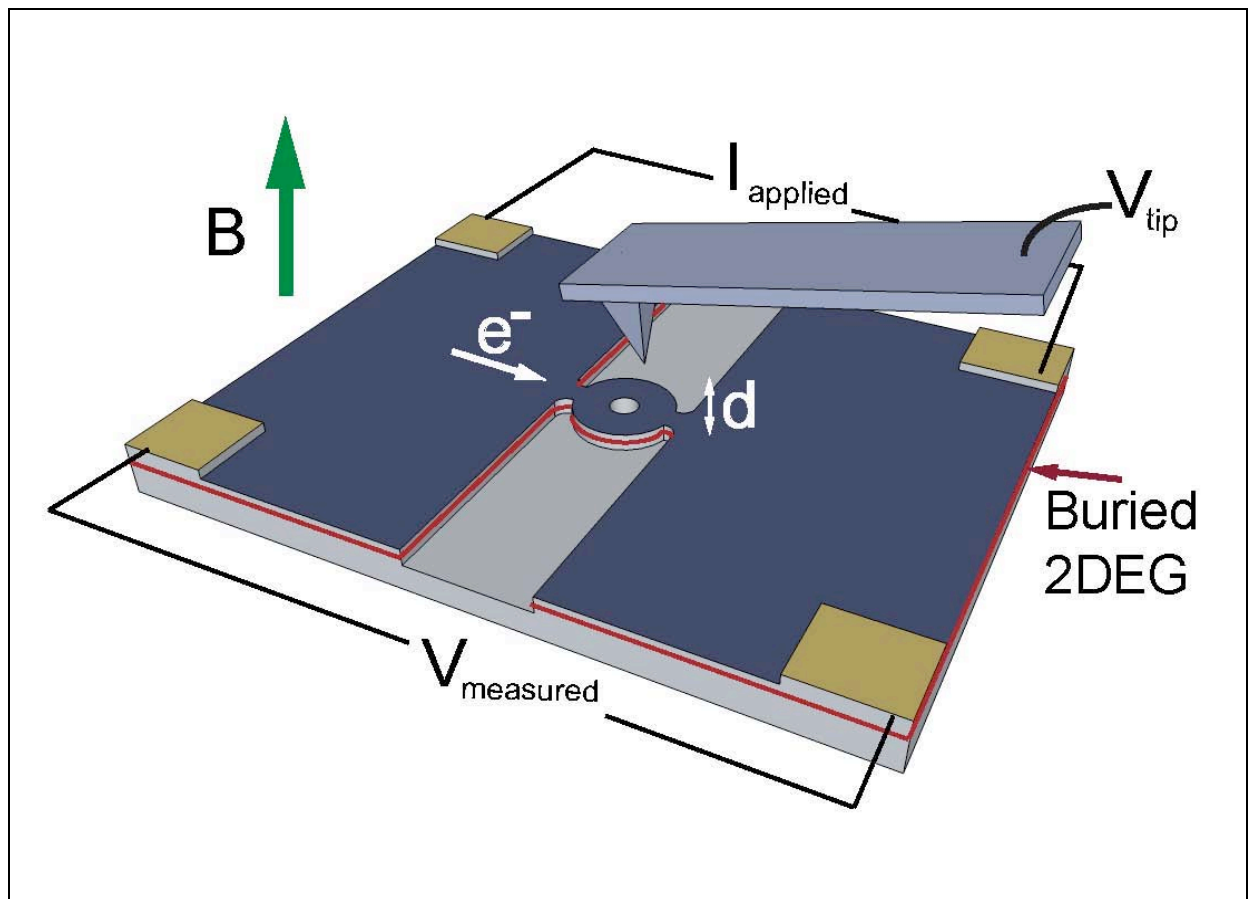

**Figure 1**. Principle of SGM. A low-frequency (about 1 kHz) probe current $I_{applied}$ (of typically 1-10 nA) is applied to the device, here a quantum ring patterned from a 2DEG buried at a few tens of nanometres below the free surface. The measurement of the voltage drop $V_{measured}$ across the device gives access to its electrical conductance $G$ or resistance $R$. The AFM tip is biased with a voltage $V_{tip}$ typically of the order of a few tenths of volts up to a few volts and scanned at a distance $d$ of typically 20 to 50 nm above the device (or alternatively, the sample is scanned under the fixed tip). This modifies the conductance (resistance). Imaging the conductance changes $\Delta G(x,y)$ or, alternatively, resistance changes $\Delta R(x,y)$ due to the tip perturbation as function of the lateral tip position $(x,y)$ builds what is called a conductance, respectively resistance, SGM image in this paper.

## 2. The technique of scanning-gate microscopy

*2.1 Principle*

The principle of SGM as applied to the imaging of a semiconductor nanostructure (here ring-shaped) patterned from a buried two-dimensional electron gas (2DEG) is sketched in figure 1. A voltage-biased ($V_{tip}$) AFM tip is scanned at an altitude of 20 to 50 nm over the device surface to perturb locally one of its macroscopic electrical properties whose induced changes are mapped as function of the tip position. In the example of figure 1, it is the conductance $G$ (or resistance) of the device that is imaged as function of the tip position in a current-biased device ($I_{applied}$) configuration. Alternatively, a voltage-biased device configuration can be used in particular in high impedance devices like quantum point contacts (QPCs) (see eg. [2,3]).

The whole setup is plunged into a cryostat to operate at low temperature, down to below 100 mK in a dilution refrigerator for the coldest SGM setups [4,5]. Optionally, an external magnetic field can be applied. The combined low-temperature and magnetic-field environment requires the use of cryogenic magnetic-free displacement units, such as for example titanium-made inertial step motors [6] for the *in situ* coarse positioning of the tip relative to the nanostructure over a few millimetres. Fine positioning over a few micrometres for the image acquisition is ensured by commercially available piezoelectric scanner elements.

*2.2 Some technical issues*

SGM experiments have several difficult steps to deal with. The most critical one perhaps is to locate the region of interest, without damage (either electrical or mechanical) to both the tip and sample, in a top-loading microscope that has no optical access to the user. This can be achieved by using the microscope in the AFM "tapping" mode to measure the topography of the sample. The surface can be supplied with a set of patterned arrows and other "traffic signs" pointing to the nanostructure, and locating these signs by recording successive images can serve as a guide to drive the tip towards the active device for subsequent SGM imaging.

Measuring the sample topography requires measuring the AFM cantilever deflection by some means. Instead of using an optical method as commonly done in AFM, it is recommended using a light-free setup such as soft piezoresistive cantilevers [1-3] or commercially available quartz tuning forks [7]. This is because semiconductor materials are photosensitive and illuminating them at low temperature can induce charge redistributions in their electrostatic background and undesired persistent changes in their carrier density [8].

Tuning forks have originally been introduced for near-field scanning optical microscopy [6,7,9] and extended later to other scanning-probe microscopy (SPM), such as for example the low-temperature combined AFM-STM [10] used to study partially conducting hybrid devices. For the purpose of SGM, either a metallic tip or a commercial conductive AFM cantilever can be electrically anchored on one metallic pad of the tuning fork. A further advantage of using a tuning fork is that both the topographic profile and conductance images can be recorded simultaneously if required. A drawback however is that the tuning fork is much stiffer than a standard AFM cantilever (static spring constant in excess of 20.000 N/m compared to a few N/m for a cantilever), and special care must be taken not to apply a too large force [7] on the sample. This can be achieved by using a phase-locked loop regulation of the tuning fork vibration. Otherwise, it could irreversibly damage the sample or rearrange residual charges in the device environment, thereby altering its electronic properties.

Once the tip has located the device, it is lifted at some tens of nanometres above the surface and scanned in a plane parallel to the 2DEG (the tip in general does not follow the topography). Therefore, all subsequent SGM measurements are carried out without contact to the surface. However, it may happen that drifts occur during SGM imaging, especially if parameters like temperature or magnetic field are varied. In this case, additional topographic images are acquired to compare with the initial topography and possibly adjust the tip position.

## 3. Some successful achievements of scanning-gate microscopy

Here, we give a summary of a few *selected* achievements of SGM, which, we believe, have given strong impulse to this imaging technique. We refer the reader to the cited literature for more information.

*3.1 Quantum-point contacts in two-dimensional semiconductor heterostructures*

One spectacular achievement is the imaging of coherent electron flow from QPCs patterned in high-mobility GaAs/GaAlAs 2DEGs. In QPCs, the conductance is quantized in units of *2e²/h* (*e* is the electron charge, *h* the Planck's constant), each plateau corresponding to a conductance channel propagating through the QPC with near unity transmittance [11]. SGM has been able to image the three lowest conductance modes in real space and has shown that each mode contributes to a number of spatial electron "strands". The latter could be indexed in correspondence with their parent's mode index, i.e., the lowest mode contributes one strand; the second mode contributes two strands, and so on. Each mode has a well-defined nodal structure that agrees well with the corresponding squared wavefunction [2].

In addition to this nodal structure, the electron flow images are decorated with fine undulations at half the Fermi wavelength that are oriented perpendicularly to the flow and extend to micrometres away from the QPC [2,3]. These interference patterns are due to coherent backscattering of electron waves between the tip depleted region and the QPC at short distance from the QPC, or between the tip and localized scattering centres at larger distances. Their occurrence confirms that the electron flow is imaged in the coherent regime of transport (the phase coherence length is of the order of a few $\mu$m in these systems at low temperature), whereas their spacing at half the Fermi wavelength can be used to probe the spatial profiles of the electron density in the 2DEG [12]. A similar electron interference imaging experiment has been done using an additional mirror gate to backscatter part of the electron flow towards the QPC [13]. Another interesting feature of the coherent flow from a QPC is that it splits in an increasing number of smaller ramifications at increasing distances from the QPC. This branching is explained as due to focusing in the valleys of the background potential and to the presence of hard scattering centres [3]. Therefore, SGM turns out to provide additionally a visual insight into this background potential.

More recent work has revealed that the larger the 2DEG mobility, the farther from the QPC does the above branching occur [14], in agreement with an increased mean free path. In the highest mobility samples at an usual operation temperature of 1.7 K, the fine decoration at half the Fermi wavelength mentioned above does not survive at large distances from the QPC as a consequence of a cleaner background potential landscape and a finite thermal smearing of the electron interferences. However, these fringes are recovered at lower temperatures (350 mK) and allow probing spatially phase coherent properties in the "clean" electron interferometer formed by the SGM tip and the QPC [15]. In a very recent work, electron-electron scattering in the 2DEG has also been studied in these interferometers [16].

The above imaging of QPCs has been extended along several directions. One is dealing with "magnetic steering". In presence of a perpendicular magnetic field, the Lorentz force bends the electron flow in the 2DEG that forms closed cyclotron orbits. SGM has revealed these cyclotron trajectories in a sample containing two QPCs in series [17,18]. Another extension deals with the visualization of quantum interference patterns within a QPC whose area distribution could be shown by SGM to be similar to that of the magneto-conductance fluctuations of the QPC, thereby indicating a common origin for both effects [19]. The last extension we wish to mention goes beyond imaging, in that the SGM tip has been used to finely tune the conductance quantization of the QPC, in particular to investigate the nature of the anomalous 0.7 plateau that often appears below the conductance quantum [20,21].

The imaging of coherent electron flow from QPCs has given strong confirmation that SGM is a powerful tool to image at the local scale, study and control electron transport in mesoscopic structures that are not accessible to STM. It has stimulated several theoretical analyses based on Green's function techniques or perturbation theory [22-25] and has led to an increasing number of experimental groups developing the technique and applying it to a broadening range of buried structures and devices as written in the remaining of this section.

*3.2 Semiconductor quantum dots and rings*

Since the SGM tip behaves as a local gate, it can be used to study Coulomb blockade in quantum dots (QDs). Scanning the biased tip around a dot draws a set of concentric isopotential lines corresponding to successive charging states of the dot. This effect allows one to precisely locate the position of a dot when it is unknown, like in the case of multiple dots along nanowires [26] and nanotubes [27], or traps in the barrier of a 2DEG heterostructure [28-30]. Coulomb blockade has also been used to investigate the spatial extension of a localized wavefunction inside a single-electron QD [31] or, conversely, to map the spatial distribution of the tip induced potential [32,33].

SGM was also applied to large open QDs, called quantum billiards, to image scarred wavefunctions corresponding to classical orbits and their periodicity in magnetic field [34]. SGM has also proven to be successful in imaging quantum rings (QRs), where it has been able to give images of Aharonov-Bohm interferences [35], of the electronic local density of states [36-38], and of Coulomb islands in the Quantum Hall regime [5]. These results on QRs are the focus of two subsections below.

*3.3 Carbon-based micro- and nanostructures*

If SGM has been primarily applied to modulation-doped structures made of conventional semiconductors, it has also been very successful with other systems such as carbon-based structures. Carbon nanotubes, for example, have been the focus of one of the very first SGM experiments that revealed the exact position of several quantum dots appearing along the nanotube and giving rise to complex Coulomb blockade conductance oscillations in transport experiments [27].

This widening interest has been confirmed by the recent publication of several papers dealing with the SGM imaging of single-layer graphene flakes [39], the study of the influence of a tip-induced scattering potential on universal conductance fluctuations [40] and on weak localization [41] in graphene devices and the imaging of graphene QDs coupled to source and drain by narrow constrictions [42]. In the latter case for instance, SGM conductance images draw three overlapping sets of concentric Coulomb rings that are centred on the dot, for one of them, and on the two constrictions, for the two others. These ring-like patterns build real-space images of the Coulomb blockade in the dot and the two constrictions, and reveal the exact location of the Coulomb-blocking species, in pretty much the same way as in the semiconductor QDs summarized in the previous subsection. In addition to this imaging ability, SGM is able to extract valuable quantitative information such as the extension of the involved localized states and the charging energy of the dot.

*3.4 Quantum rings in the Aharonov-Bohm regime*

An open QR in the coherent regime of transport sees its conductance increasing to a maximum when electron waves interfere constructively at the output contact and decreasing to a minimum for destructive interferences. Varying either the magnetic flux captured by the QR or the electrostatic potential in one arm allows the interference to be tuned by changing the phase accumulated by electrons as they are transmitted through the QR's arms. This gives rise to the well-known magnetic and electrostatic [43,44] AB oscillations in the ring conductance [45]. An example of AB oscillations is given in figure 4a, showing the magnetoresistance of a QR with a radius of 520 nm, and a temperature of 90 mK. The temperature dependence of the AB oscillations amplitude is directly related to that of the phase coherence length or time, and saturates at low temperature, when the phase coherence time is equal to the dwell time (i.e., the time spent by electrons inside the device, statistically), as illustrated in figure 1 of [36]. Since the dwell time depends on the QR area and on the number of modes in the QR openings, this saturation occurs at different temperatures for different quantum ring geometries [46].

SGM can image these archetypal interference phenomena in real space [35]. The electrostatic AB effect gives rise at low magnetic field to a well-developed fringe pattern in the conductance image of GaInAs-based QRs in the coherent regime of transport when the tip scans outside the QR, as illustrated in [35]. This outer pattern is mainly concentric with the ring geometry. The qualitative interpretation (see [47] for a more quantitative approach) in terms of a scanning-gate-induced electrostatic AB effect is that as the tip approaches the QR, either from the left or right, the electrical potential mainly increases on the corresponding side of the QR. This induces a phase difference between electron wavefunctions travelling through the two arms of the ring, and/or bends the electron trajectories, which produces the observed pattern. Modifying the magnetic field strength contributes another phase term through the magnetic AB effect and displaces the whole pattern with respect to the QR. This displacement is periodic in magnetic field strength with the same periodicity as the AB oscillations seen in the magneto-conductance, which gives further support to the interpretation in terms of AB effects.

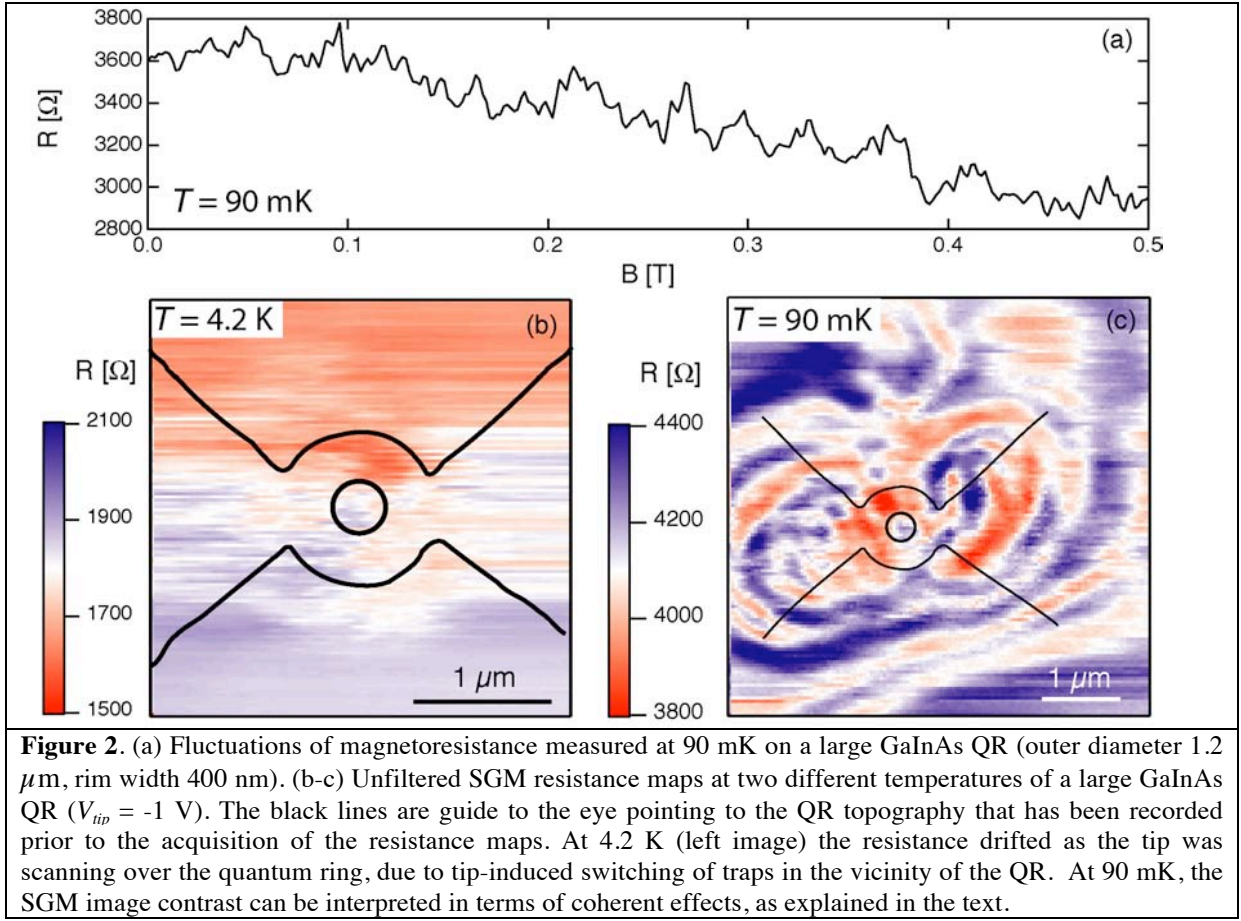

**Figure 2**. (a) Fluctuations of magnetoresistance measured at 90 mK on a large GaInAs QR (outer diameter 1.2 $\mu$m, rim width 400 nm). (b-c) Unfiltered SGM resistance maps at two different temperatures of a large GaInAs QR ($V_{tip}$ = -1 V). The black lines are guide to the eye pointing to the QR topography that has been recorded prior to the acquisition of the resistance maps. At 4.2 K (left image) the resistance drifted as the tip was scanning over the quantum ring, due to tip-induced switching of traps in the vicinity of the QR. At 90 mK, the SGM image contrast can be interpreted in terms of coherent effects, as explained in the text.

In this framework, one can expect that the evolution of the SGM contrast with temperature be related to the temperature dependence of the AB effect. The confirmation of this hypothesis is shown in [36] (figure 1), for the range 3-40 K. Note that in the case of this particular QR, both the SGM contrast and the AB oscillation amplitude saturate below 10 K, an effect related to the dwell time, as explained above. In QRs with different geometrical parameters, it can be advantageous to go to lower temperatures to enhance the SGM contrast, as illustrated in figure 2. Here, the QR has both a larger diameter (1.2 $\mu$m) and wider arms (400 nm) than those studied in our previous works [35,36]. In this geometry, electrons have the opportunity to probe a much larger set of semiclassical trajectories. As a consequence, the signature of electron interference in the magnetoresistance is no longer made of oscillations with a single characteristic frequency, but rather aperiodic reproducible fluctuations, the so-called "Universal Conductance Fluctuations", shown in figure 2a. The superposition of trajectories also induces an "averaging" effect, which reduces the amplitude of the contribution of coherent effects in the magnetoresistance, compared to rings with narrower arms. As a consequence, it is necessary to reach lower temperatures to recover a good contrast in SGM images. This is shown in figures 2b and 2c, comparing SGM resistance images measured on this "large" QR at two temperatures, T= 4.2 K and 90 mK (and zero magnetic field). Note that the images of figure 2 are unfiltered original images in contrast with [35], where a slowly varying background tip effect was subtracted. The amplitude of the contrast related to coherent effects in SGM images is strongly enhanced at the lowest temperature, and a rich set of details is revealed. These data therefore clearly show that the improvement of SGM when the temperature decreases depends on the system under investigation.

QR conductance images also exhibit a complex pattern when the tip scans directly over the QR region (see figure 4d and data in [35-38]). These inner fringes have been discussed in our previous papers and linked to the electron-probability density in the QR [36-38]. A detailed analysis based on quantum mechanical simulations of the electron probability density, including the perturbing tip potential, the magnetic field, and the presence of randomly distributed impurities, is able to reproduce the main experimental features and demonstrates the relationship between conductance maps and electron probability density maps at the Fermi level. An example of such a relationship is shown in figure 3 in the case of a realistic QR perturbed by negatively charged impurities [48]. Although the impurities distort the electronic local density-of-states (LDOS), this distortion is reflected back in the conductance image in such a way that the conductance map can still be seen as a mirror of the LDOS, provided that the perturbation induced by the tip is "small and narrow enough" (details concerning the range of validity of the correspondance LDOS-SGM map in QRs can be found in [36-38]; see also subsection 4.3 below). As seen in figure 3, both the LDOS and conductance images tend to develop radial fringes, which are mostly, but not entirely, anchored to the impurity locations and are reminiscent of the experimental images that are often found to be asymmetric. Note that it is hopeless to try to find a direct and exact correspondence between simulations such as those presented in figure 3 and experimental data, unless the precise geometry of the experimental device and impurity distribution are taken into account. Indeed, the LDOS pattern in mesoscopic structures is extremely sensitive to slight changes in the electron confining potential (as shown in [36]). However, qualitative and general trends, such as the tendency to develop radial fringes, can be compared.

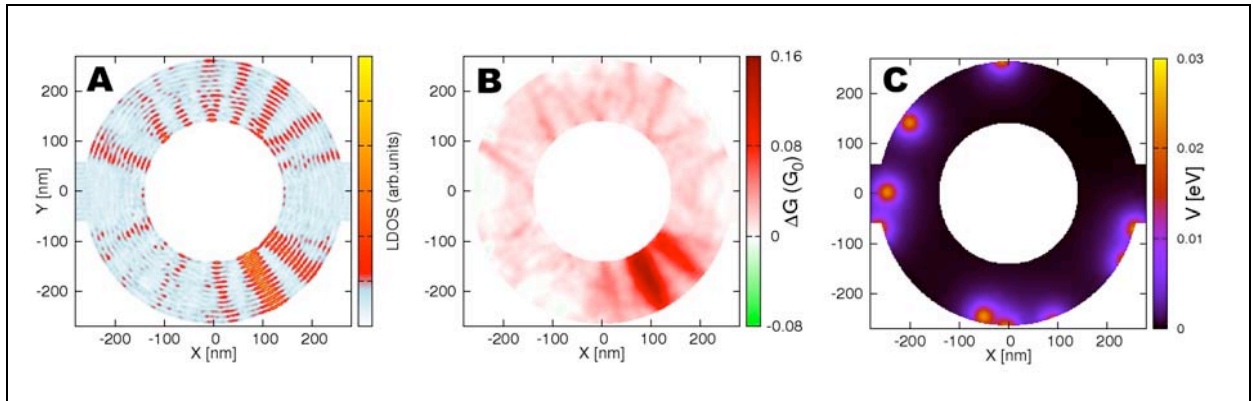

**Figure 3.** Quantum mechanical simulation of a SGM experiment on a QR in the presence of negatively charged impurities. The outer diameter, inner diameter and opening width are 530 nm, 280 nm, and 120 nm, respectively. The effective mass is 0.04 $m_0$, with $m_0$ the free electron mass. The Fermi energy is $E_F$= 95.9 meV. (a) and (b) are the simulated images of the LDOS and conductance changes (in units of $G_0$, the quantum of conductance), respectively, calculated for the random distribution of negatively-charged impurities shown in (c). In the simulation, the tip potential has a Lorentzian shape with a 10 nm range and repulsive amplitude of $E_F/50$. Reprinted from [38].

The discussion above, as well as data presented in [35-38], suggest that SGM can be viewed as the analogue of STM [49] for imaging the electronic LDOS at the Fermi level in open mesoscopic systems buried under an insulating layer or the counterpart of the near-field scanning optical microscope that can image the photonic LDOS in confined nanostructures, provided that the excitation light source can be considered as point-like [50] such as in active tips based on fluorescent nano-objects [51,52]. However, imaging the LDOS of a buried electron system with SGM requires some caution to be taken, as the technique is less direct than STM applied to surface electron systems. This point is discussed further below in subsection 4.3.

*3.5 Quantum rings in the quantum Hall regime*

Several groups have applied SGM to study 2DEGs in the quantum Hall regime, for instance to image edge states, localized states, and inter-edge-state scattering centres in this high magnetic field regime

[53-55]. In this subsection, we wish to show how SGM can be useful in the study of the QHE physics by describing in some detail our study dealing with QRs in this regime.

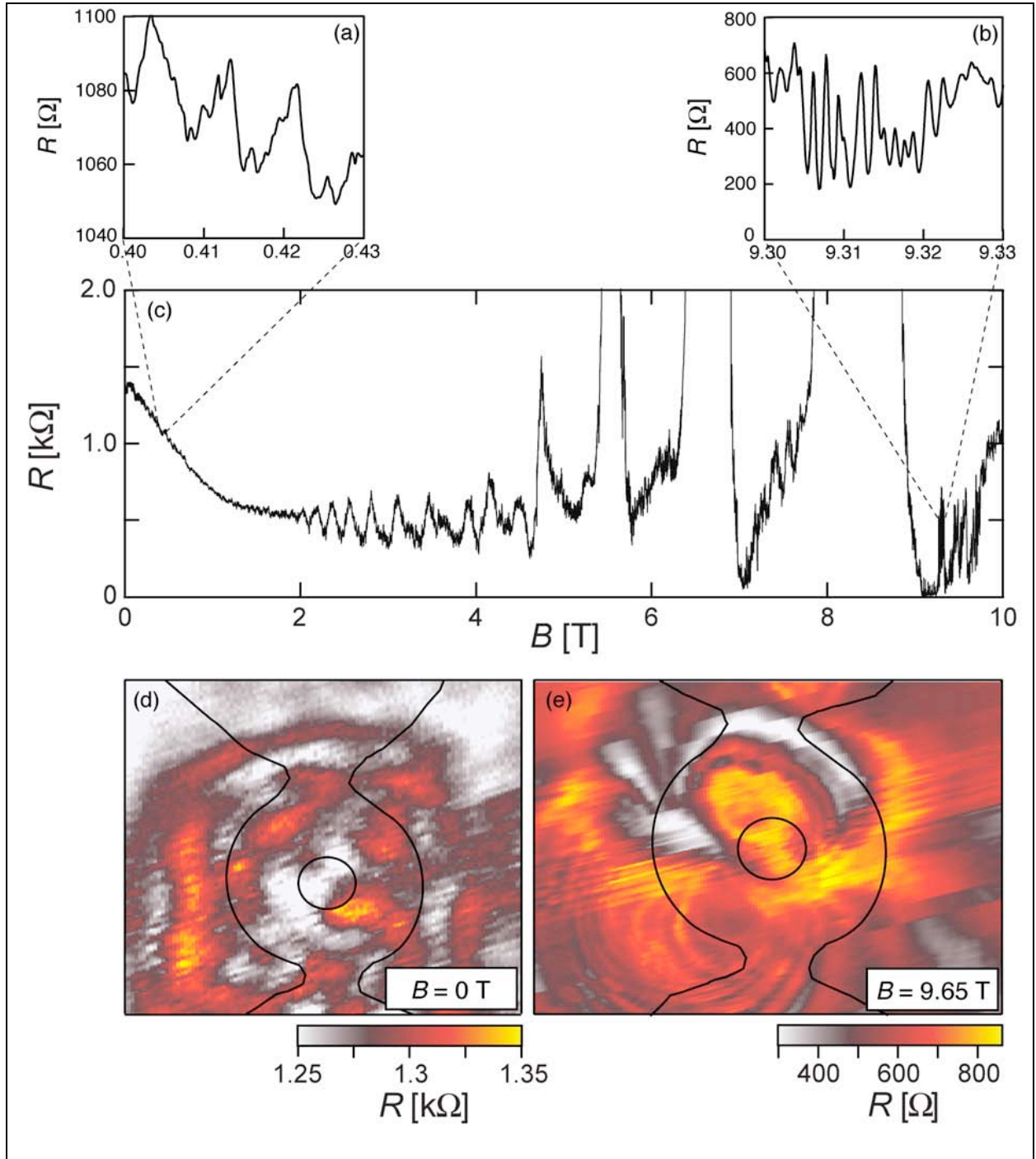

**Figure 4**. (a-c) Magnetoresistance of a QR patterned from an InGaAs/InAlAs heterostructure. The lithographic inner and outer *radii* of the SR are 215 and 520 nm, respectively, and the opening widths are 300 nm. A 2DEG is buried at 25 nm below the free surface with a carrier density of $1.4 \times 10^{12} cm^{-1}$. The curve in (c) is the QR magnetoresistance, whose low-field and high-field parts are zoomed in (a) and (b). (d-e) Two typical unfiltered SGM resistance maps taken at zero magnetic field (d) and in a high field of 9.65 T (e). The voltage applied to the tip is +0.5 V at zero field and +1 V at high field. The tip to surface distance is 50 nm. All data are taken at a temperature of 100 mK.

In QRs, the AB related phenomena described in the previous subsection occur at low magnetic field and they all disappear at higher field. This is because at high field, the cyclotron radius shrinks below the width of the QR arms and openings, and electrons tend to propagate along the edges of the device. This completely changes the electron dynamics and the contrast in SGM images, as shown hereafter.

The gradual evolution to the high-field regime is illustrated in figures 4a-c, displaying the magnetoresistance of a 1 $\mu$m-diameter QR, measured at 100 mK (same ring as in [5]). Below B = 2 T, the magnetoresistance shows periodic oscillations superimposed on a smooth background that can clearly be seen in the low-field zoom on the magnetoresistance in figure 4a. In this B-range, the SGM contrast is determined by interference effects (concentric features in the vicinity of the quantum ring), and by the details of the electron probability density in the quantum ring region, as explained above (see also [35]).

Above 2T, the ring magnetoresistance first exhibits Shubnikov-de Haas oscillations, and, at higher B, signatures of the quantum Hall (QH) regime [56], with an important difference with respect to the QH effect in macroscopic samples. Indeed, in the B-ranges where the resistance drops to zero in macroscopic Hall bars (around integer filling factors), the ring magnetoresistance shows a rich pattern of fluctuations [57-59]. An example of such fluctuations is shown in the high-field zoom on the magnetoresistance, in figure 4b. To understand the origin of these fluctuations, different models have been proposed, based either on direct tunnelling between counter-propagating QH edge states, or on tunnelling through localized states ("Coulomb islands") inside the device [60]. Such Coulomb islands can be pinned by local maxima or minima in the 2DEG potential. When the flux of magnetic field through the area of the island increases by one magnetic flux quantum, one electron must be added to each occupied Landau level in the island. This induces a change in the island charging energy, which therefore oscillates, with a period given by $\phi_0/f_c$, where $f_c$ is the number of occupied Landau levels and $\phi_0$ is the flux quantum [60].

When the biased tip of the SGM approaches such a Coulomb island, the associated modification in the potential pinning the island results in a change of the island area and, hence, of the magnetic flux through this island. As this magnetic flux controls the number of trapped electrons in the island, this induces Coulomb-type oscillations in SGM images. This is illustrated on the high-field (9.65 T) low-temperature resistance map in figure 4e. Different sets of concentric resistance fringes are visible in the image, in the lower left part of the ring, and above the central hole of the ring. The centre of each set of fringes corresponds to a localized Coulomb island, as shown in [5].

In this regime, SGM therefore allows to locate individual QH Coulomb islands, and to "manipulate" them individually, through local modifications of the electron confining potential. Knowing the location of each island, and observing the dependence of the SGM fringe pattern with the magnetic field leads to a detailed understanding of the magnetoresistance fluctuations around integer filling factors in mesoscopic devices. These experiments also open the way to the spectroscopic investigation of each QH island by changing the voltage on the tip, and the bias through the QR.

*3.6 Related scanning probe techniques for buried 2DEG investigations*

Finally, it is worth mentioning here two other cryogenic SPM techniques related to SGM, which have also proven very powerful in imaging buried electron systems, namely scanning single-electron transistor (SET) microscopy and scanning subsurface charge accumulation (SCA) microscopy.

In scanning-SET microscopy [61,62], a SET is fabricated at the apex of a tapered optical fibre and used as a scanning electric probe sensitive to any potential change in the scanned device. This technique has enabled to image the distribution of charges and the way they localize in space in GaAs/GaAlAs heterostructures under various ground state conditions including the integer [62,63] and fractional [64] QHEs. It has recently been used to image electron-hole puddles in the vicinity of the neutrality point in graphene monolayers [65,66].

In scanning SCA microscopy [67-69], the sharp metallic tip of a scanning microscope is connected to a very sensitive charge detector that records the local ac-charge accumulation in the 2DEG in response to a small ac-voltage applied to the sample. When the tip scans above a conducting region the ac-current is significant, whereas it disappears above insulating, incompressible, or Coulomb blocked regions. This imaging technique has been applied on 2DEG in the quantum Hall regime, revealing a rich structure of filaments and droplets [67], random electrostatic potential fluctuations [68], and short length-scale scattering potentials [69].

## 4. Some limitations of the SGM technique

Based on our experience in the SGM imaging of QRs, we exemplify in this section a few limitations of the technique that can be helpful to understand its capabilities and that should be considered carefully during SGM experiments.

*4.1 Tip-induced effects and artefacts in SGM*

Experimentally evaluating [33] or theoretically computing the tip potential seen by the electrons probed in SGM is a complex matter. But this is required for a proper understanding of the SGM images. For instance, the AB effects in QRs introduced earlier in this paper have indicated that the tip potential is strongly screened by the electron system [37,47].

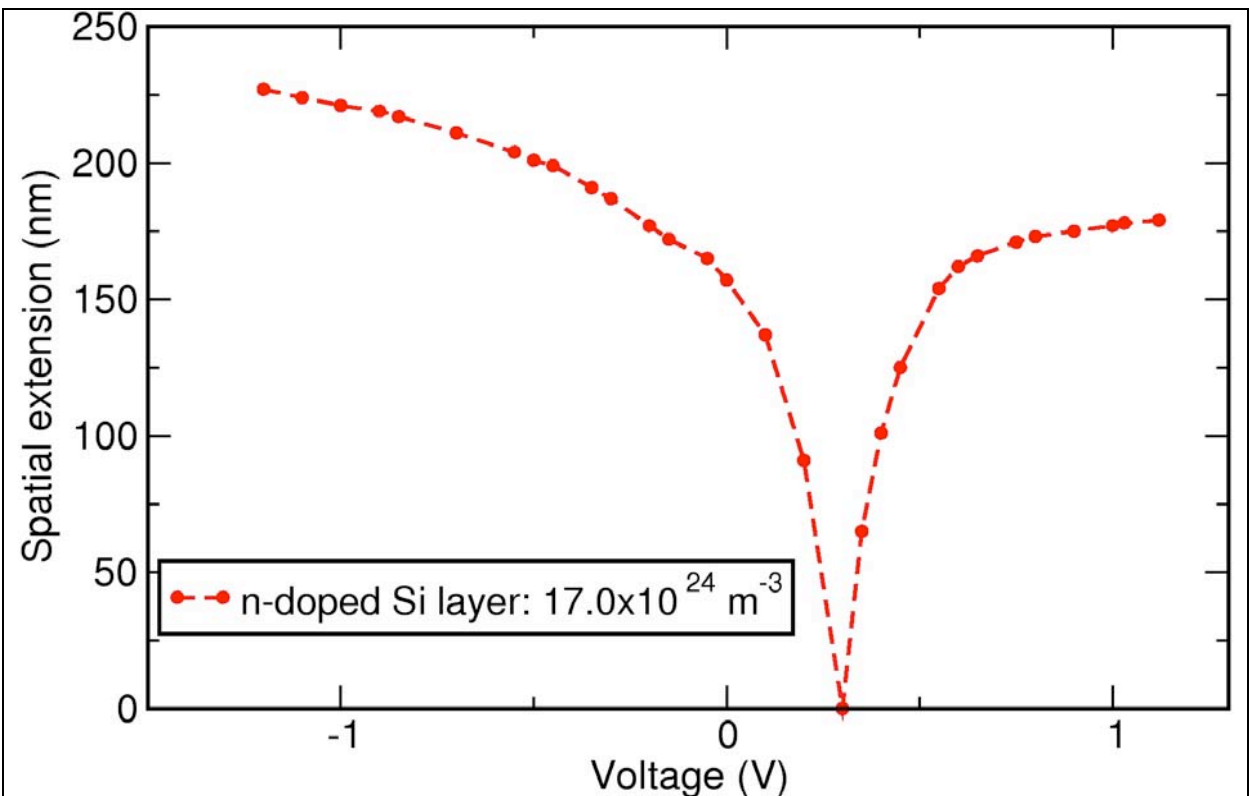

**Figure 5.** Spatial extension as a function of the tip voltage of the tip-induced potential on the 2DEG present at the interface of a GaInAs/AlInAs heterostructure with a silicon doping layer of $1.7 \times 10^{19}$ cm$^{-3}$ in the barrier.

Solving self-consistently the three-dimensional Schrödinger and Poisson equations that govern the properties of the electron system coupled to the tip potential can simulate this screening. This allows estimating the spatial extension of the tip-induced perturbation, which is a critical factor to ensure a correct interpretation of conductance images of QRs in terms of LDOS [36,37]. As an example, we show in figure 5 the dependence of the tip-induced spatial extension on the tip voltage for an SGM tip located 10 nm above the free surface of a GaInAs/AlInAs heterostructure. This spatial extension is defined as the length of the area occupied by a potential higher than a threshold value of 0.3 meV. The linear increase of the spatial extension at low tip-bias is followed by a power-law dependence due to the screening of electrons in the 2DEG. In figure 5, there is a distinctive non-zero spatial extension at zero tip voltage, or conversely a finite voltage must be applied to have a vanishing extension. This is due to the dielectric screening arising from fixed charges in the doping layer of the heterostructure and on the exposed surface. This screening is modelled as explained in [70].

On the experimental side, some caution must be paid to avoid tip-induced artefacts. A first artefact is related to possible distortions of the tip, which may happen after thorough scanning of the sample or inadvertent contact to the surface. It results in distortions of the SGM images, which can also be simulated (see figure 2 in [37]). For large distortions, it may even happen that the respective roles of the electron system and of the tip are reversed, as suggested by the SGM image in figure 6b. Here the tip has been damaged during scanning of the sample in such a way that the tip potential is imaged rather than the intrinsic properties of the electron system. Fortunately, a strong indication that the tip has been damaged and that the SGM data are consequently not reliable is given by the strongly distorted topographic image in figure 6a, where the ring geometry can hardly be recognized in contrast with topographic images acquired with good tips (see e.g. figure 1a in [35]).

It may also happen that the tip is contaminated by particles that distort the induced potential. In this case a high-field tip treatment could be used to recover a smooth and symmetrical potential: a high voltage pulse is applied on the SGM tip in STM mode on a metallic electrode to eject the unwanted particles before SGM imaging of the nanostructure [71].

*4.2 Distortions due to random residual charges*

SGM images can be disturbed by the presence of parasitic electron traps coupled by tunnelling to the 2DEG of the semiconductor device under investigation. These traps may originate from impurities located in the spacer layer of the heterostructure very close to the 2DEG, closer than the doping layer whose charge transfer is blocked at low temperature. The time fluctuation of these electron traps is a frequent source of low-frequency noise in high mobility transistors. When the SGM tip scans over such an impurity, its charge state can change, and consequently, the potential landscape for the 2DEG electrons also changes, thereby affecting the conductance of the device. At low temperature, these traps are subjected to Coulomb blockade, and the number of electrons changes one by one in a discrete manner when the tip is approached. Since the charge state is the same when the tip is at a constant distance from the trap, SGM images display sets of concentric rings centred on each trap coupled to the 2DEG. The observed pattern changes with illumination (that acts as a charge reset) or after a new cooling down of the sample. This parasitic effect in SGM has been observed for instance in QPC experiments (see figure 3 in [28]) and in QD studies (see figure 2c in [29]).

*4.3 On the imaging of the LDOS of a buried electron system*

STM in its spectroscopic mode – scanning tunnelling spectroscopy - is known to probe directly the electronic LDOS of a surface electron system, such as for example the quantum corral [49] or the inversion layer on an InSb surface in the QHE regime [72], which can be computed accurately [73]. Since SGM probes tip-induced changes in the conductance maps of a buried 2DEG, the connection of the collected information with the LDOS is indirect and some caution must be taken in interpreting the images. We have already mentioned this issue earlier in this paper (subsection 3.2). Here we wish to put more emphasis on this important point in the case of QRs.

According to the analytical model presented in [37], in the single-channel transmission case a generalized Kramers–Kronig relation holds between the LDOS and the conductance variation due to the tip-induced potential. This is valid in the linear response regime where the local perturbation due to the tip is so weak that the energy levels of the system are unchanged. In the multi-channel transmission case, an exact correspondence between LDOS and conductance maps is not possible due to the deficient spatial resolution of the tip-induced potential and to the uneven effect of the tip on the different transverse modes involved in the transport.

However, a noticeable correspondence is still achievable in some cases of interest for which the total wavefunction is determined only by the few states close to the Fermi energy and presents a large and uniform spatial extension. Examples are given by mesoscopic systems such as QRs where the LDOS is dominated by localized states due to randomly distributed impurities or by semi-classical periodic orbits inducing scarring effects in the total wavefunction.

In addition to these fundamental limitations, special care must be taken to apply a small bias to the tip. Indeed, applying too high a voltage will obviously strongly couple the tip to the buried electron system and eventually modify its properties. This can be simulated without resorting to self-consistent simulations as done in [36] where it was found that for small tip potentials the conductance image essentially reflects the electronic LDOS in the QRs. However, applying too high a tip potential progressively introduces spurious features in the conductance image that are not present in the LDOS (see figure 4 in [36]). Experimentally, the low tip-potential regime is maintained as long as the magnitude of the imaged tip-induced features, such as the AB "inner fringes" mentioned earlier, increases linearly with the tip bias [36]. Beyond this regime, this magnitude tends to increase sub-linearly, or to saturate, and spurious features, not seen at low bias, appear. This regime must be avoided. Note however that this limitation of a low bias voltage applied to the tip holds essentially for the SGM experiments aiming at measuring the unperturbed electronic properties of the electron system such as the LDOS. It does not hold for measurements where the tip is used to scatter the electron waves as done, for instance, in the imaging of coherent electron flow through QPCs or related experiments [2,3,12-16].

## 5. Conclusion

The few examples summarized in this paper confirm that SGM is very powerful in imaging the electronic transport in various low-dimensional semiconductor devices and to reveal how electrons behave down there. It often gives valuable complementary view on phenomena that are usually considered within a macroscopic experimental scheme. The ability of locating precisely compressible Coulomb islands in a quantum Hall interferometer is illustrative of this claim. Although some attention must be paid to avoid possible artefacts, the broad applicability range of SGM makes it a powerful tool for the electron diagnose of nanodevices in the coherent regime of transport, or even in the quantum Hall regime. Therefore, more spectacular achievements can be expected in the future.

**Acknowledgments**

B. H. is FNRS research associate and F. M. is FNRS postdoctoral researcher. F. M. acknowledges a former postdoctoral grant with FCT - Portugal. This work has been supported by FRFC grant no. 2.4.546.08.F, and FNRS grant no 1.5.044.07.F, by the Belgian Science Policy (Interuniversity Attraction Pole Program IAP-6/42) as well as by the PNANO 2007 program of the Agence Nationale de la Recherche, France ("MICATEC" project). VB acknowledges the award of a "Chaire d'excellence" by the Nanoscience Foundation in Grenoble.